\documentclass[%
 10pt,twoside,twocolumn,english,aps,amssymb,amsmath,amsfonts,superscriptaddress,showpacs,prb
]{revtex4-1} 

\usepackage{graphicx}
\usepackage{dcolumn}
\usepackage{bm}
\usepackage[T1]{fontenc}
\usepackage{natbib} 
\usepackage[latin9]{inputenc}
\usepackage[active]{srcltx}
\usepackage{babel}
\usepackage{array}
\usepackage{mathrsfs,ulem,epstopdf}
\usepackage{bm}
\usepackage{wasysym}
\usepackage{color}
\usepackage[colorlinks,bookmarks=false,citecolor=blue,linkcolor=blue,urlcolor=blue]{hyperref}
\def\wn/{\,cm$^{-1}$}
\def\area/{\,cm$^{-2}$}
\def\cubic/{$_\mathrm{c}$}
\def\DM/{Dzyaloshinskii-Moriya}
\def\BTCP/{Ba(TiO)Cu$_4$(PO$_4$)$_4$}
\DeclareRobustCommand{\rchi}{{\mathpalette\irchi\relax}}
\newcommand{\irchi}[2]{\raisebox{\depth}{$#1\chi$}} 

\makeatletter
\@ifundefined{textcolor}{}
{%
 \definecolor{BLACK}{gray}{0}
 \definecolor{WHITE}{gray}{1}
 \definecolor{RED}{rgb}{1,0,0}
 \definecolor{GREEN}{rgb}{0,1,0}
 \definecolor{BLUE}{rgb}{0,0,1}
 \definecolor{CYAN}{cmyk}{1,0,0,0}
 \definecolor{MAGENTA}{cmyk}{0,1,0,0}
 \definecolor{YELLOW}{cmyk}{0,0,1,0}
}

\makeatother

\begin{document}


\title[Sample title]{Magnetic structure in square cupola compound Ba(TiO)Cu$_4$(PO$_4$)$_4$: a $^{31}$P NMR Study}

\author{Riho R\"{a}sta}
\author{Ivo Heinmaa}
\affiliation{National Institute of Chemical Physics and Biophysics, Akadeemia tee 23, 12618 Tallinn, Estonia}
\author{Kenta Kimura}
\author{Tsuyoshi Kimura}
\affiliation{Department of Advanced Materials Science, University of Tokyo, Kashiwa, Chiba 277-8561, Japan}
\author{Raivo Stern} 
\affiliation{National Institute of Chemical Physics and Biophysics, Akadeemia tee 23, 12618 Tallinn, Estonia}

\date{\today }
\begin{abstract}
	The magnetic structure of the antiferromagnetic square cupola compound \BTCP/ with the tetragonal structure is studied with $^{31}$P nuclear magnetic resonance techniques. The magnetic hyperfine shift $K $ shows a clear splitting at the N\'{e}el temperature T$_N$~=~9.5~K, where the resonance splits into two lines when an external magnetic field is oriented along the $c$ axis and into four lines when the field is along the $a$ axis. In the paramagnetic region $K(T)$ follows temperature dependence of the magnetic susceptibility $\chi(T)$. From $K$ vs $\chi$ plot we determined nearly isotropic hyperfine field values \textit{H}$_{\text{hf}}^{a}$ = 765 mT/$\mu_B$ and \textit{H}$_{\text{hf}}^{c}$ = 740 mT/$\mu_B$ for the magnetic field oriented along $a$ and $c$, respectively. From the rotation of the single crystal in the external magnetic field we determined eight different orientations of $K$-tensor in the paramagnetic region. In the antiferromagnetic state at $T$~=~6~K we found that the local field at phosphorus is mainly due to dipolar field of coppers. Here the rotation of the single crystal shows eight different orientations of the local field $B_\text{int}~=~35.6$~mT. The orientations correspond to the calculation of dipolar fields at phosphorus assuming magnetic quadrupolar configuration of magnetic moments $\varGamma_3(1)$ described previously [\textit{Nat. Commun.} \textbf{7}, 13039 (2016); \textit{Phys. Rev. B} \textbf{96}, 214436 (2017)].	
\end{abstract}


\maketitle

\section{Introduction}
\setlength{\parskip}{0em}
\begin{figure*}
		\includegraphics[width=1.0\textwidth]{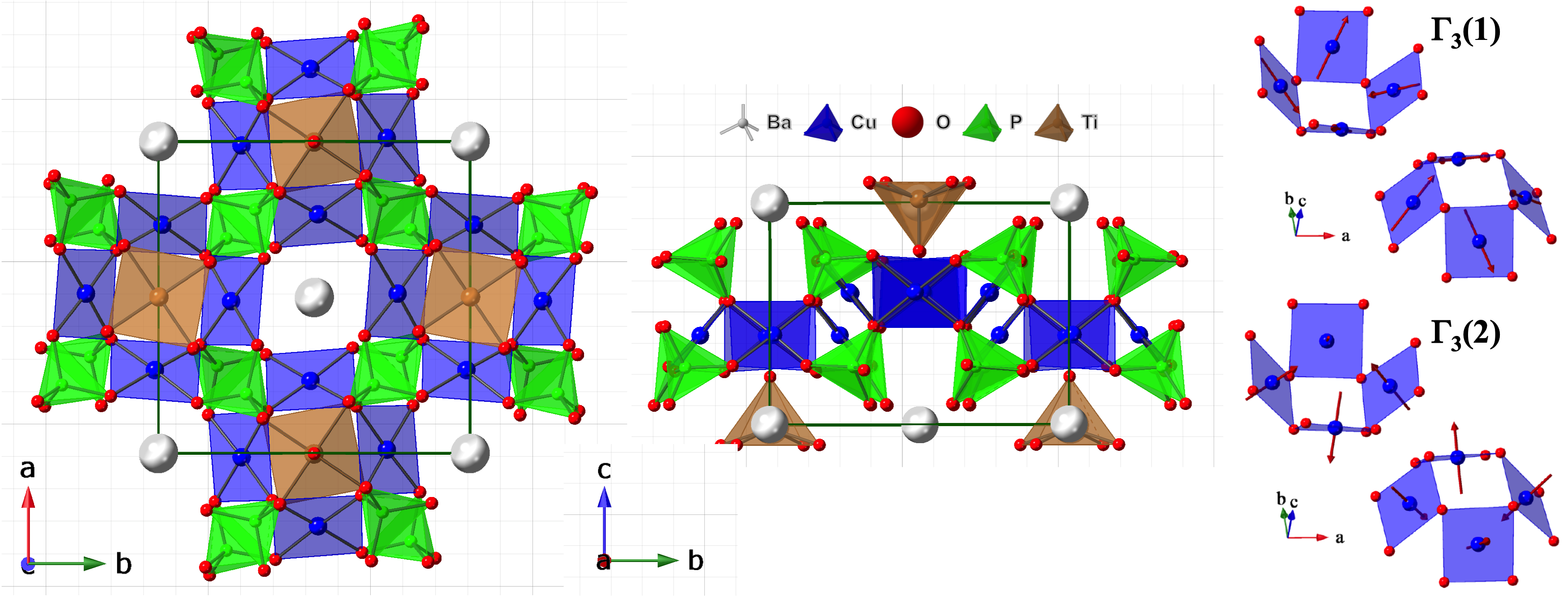} 
		\caption{\label{struc} Description of BTCPO unit cell with up and down Cu$_{4}$O$_{12}$ square cupola (blue), PO$_{4}$ tetrahedra (green), and TiO$_{5}$ pyramids (brown), and Ba ions as large grey spheres along the $c$ (left panel) and $a$ (middle panel) axis. Two possible arrangements of the spins $\varGamma_3(1)$ and $\varGamma_3(2)$ (right panel).}
\end{figure*}

Recently, an interesting class of compounds,$AE$(TiO)Cu$_4$(PO$_4$)$_4$ ($AE$TCPO; $AE$ = Ba, Sr, or Pb), was newly synthesized~\cite{Kimura2016magneto,Kimura2018a} . 
The structure of $AE$TCPO is tetragonal with the space group $P$42$_1$2 and consists of the layers of up and down square cupola Cu$_{4}$O$_{12}$  (FIG.~\ref{struc}). Each square cupola is made of four corner sharing CuO$_{4}$ plaquets. The layers are separated by $AE$ ions and TiO$_{5}$ pyramids. 
These compounds undergo an antiferromagnetic (AF) transition at low temperatures. For example,  \BTCP/ (BaTCPO) exhibits an AF ordered  state at temperatures below the N\'{e}el temperature T$_N$~=~9.5~K. Below T$_N$
the arrangement of magnetic moments of coppers shows an antiferroic quadrupolar order. Because of the antiferroic order, BaTCPO does not exhibit a magnetoelectric effect where an electric (magnetic) field causes magnetization (electric polarization). 
Instead, this compound exhibits a remarkable magnetodielectric effect. Sr(TiO)Cu$_4$(PO$_4$)$_4$ (SrTCPO) also shows similar magnetic and magnetodielectric properties below T$_N$~=~6.2~K~\cite{SrPRRB2019}, while Pb(TiO)Cu$_4$(PO$_4$)$_4$ (PbTCPO) exhibits a ferroic quadrupolar order resulting in a linear magnetoelectric effect  below  T$_N$~=~6.5~K ~\cite{PbPRMat2018,PbJPSJ2019}.

Magnetic structure of BaTCPO was carefully studied by neutron scattering experiments ~\cite{Kimura2016magneto}. 
These studies yielded two possible arrangements of the spins: one, denoted $\varGamma_3(1)$, where the moments are confined approximately in the CuO$_{4}$ planes; and the other, $\varGamma_3(2)$, where the moments are approximately perpendicular to the CuO$_{4}$ planes forming two-in-two-out-type structure (FIG.~\ref{struc}). The latter arrangement had smaller R-factor, thus being the most probable arrangement. Later Babkevich \textit{et al.}~\cite{Babkevich2017} confirmed the two possible structures using spherical neutron polarimetry, inclining by a discernible, albeit small advantage towards the $\varGamma_3(2)$ spin structure. 

In the present report we will use $^{31}$P NMR techniques to study the local magnetic fields in a single crystal of BaTCPO. Phosphorus ions in PO$_{4}$ tetrahedra see four coppers in two cupolas as next-nearest-neighbours. Therefore, the local field at phosphorus is definitely influenced by the magnetic arrangement of coppers. Previous NMR studies of SrTPCO~\cite{Islam2018} and BaTPCO~\cite{Kumar2019} have used powder samples where detailed information about the magnetic structure is difficult to get.

\section{Experimental details}
Single crystals of \BTCP/ were grown with the flux method by Kimura \textit{et al.}~\cite{Kimura2018a,Kimura2016inorganic}. The sample crystal used in the experiments sized 1.9~x~2.0~x~3.9~mm and weighed 61.84~mg. NMR measurements were conducted using the spectrometer MAGRes2000 attached to a B = 4.7~T superconducting magnet, $^{31}$P resonance frequency 80.97~MHz. He-flow cryostat (JANIS Research Inc.) allowed measurements in the temperature range of 4.5 to 300~K. The NMR probe was our own made and had a single-axis goniometer. LakeShore CERNOX  calibrated sensors and  Model 332 temperature controller were used for temperature regulation. Spin-lattice relaxation T$_1$ was measured using inversion-recovery method. The magic angle spinning (MAS) room temperature spectrum was recorded using a home-built probe with the $1.8$~mm O.D. rotor spinning $25$~kHz. The magnetic shifts are given relative to the resonance of H$_3$PO$_4$. 

\section{Results and analysis}
\subsection{Magnetization}
\begin{figure}[!]
	    \centering
		\includegraphics[width=0.5\textwidth]{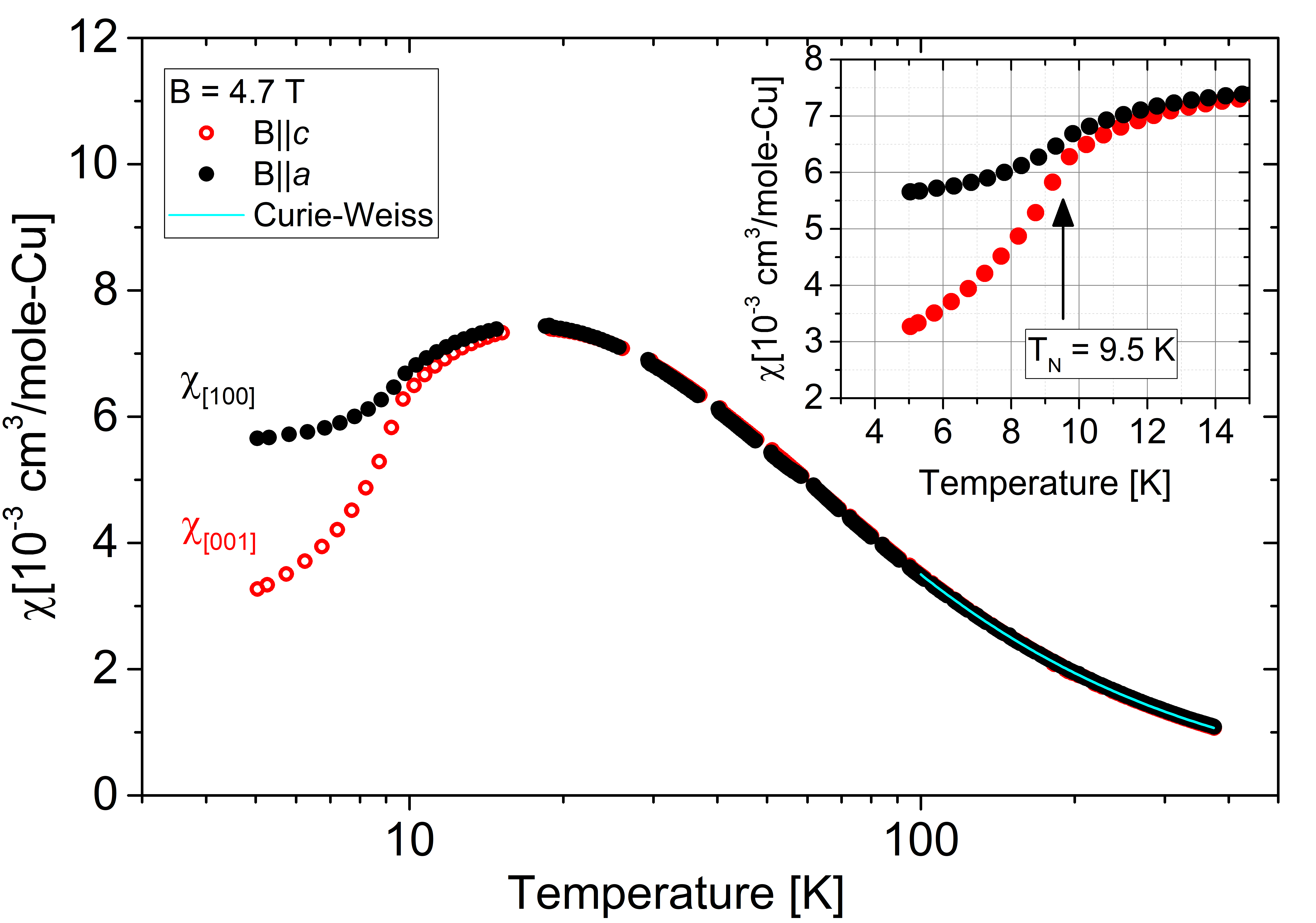}
		\caption{\label{susc}Temperature dependence of magnetic susceptibility $\rchi(T)$ in an applied magnetic field B~=~4.7~T, measured in two different directions: B$\parallel$[100] (black, filled) and B$\parallel$[001] (red, empty). The cyan line represents the Curie-Weiss fit. The inset shows a detailed behavior of susceptibilities below T$_N~=~9.5$~K.}
\end{figure}

The temperature dependence of magnetic susceptibility was measured with the applied magnetic field B = 4.7~T along the [001] and [100] directions as shown in FIG.~\ref{susc}. At temperature region T$^{\text{max}}\approx$~17~K there is a broad maximum of $\rchi$ vs T curve, which indicates onset of short range order.

At high temperature $\rchi(T)$ followed the Curie-Weiss law:
\begin{align}
	\rchi(T)=\rchi_0+\frac{C}{T-\theta_\text{CW}}.
\end{align}
Here $\rchi_0$ is temperature-independent susceptibility. 
At T~$ > $~100~K the fitting gives the following parameters: for the magnetic field along the [001] direction: 
$\rchi_{0}~=~-7.3(4) \times 10^{-5}$~cm$^3$/mole per Cu, 
$C~=~0.460(1)$~cm$^3$~K  per mole Cu, 
$\theta_{\text{CW}}~=~-29.0(4)$~K; and for the field along the [100] direction: 
$\rchi_{0}~=~-3.5(2) \times 10^{-5}$~cm$^3/$mole-Cu, 
$C~=~0.454(7)~\times10^{-4}$~cm$^3$~K $/$mole-Cu, 
and $\theta_{\text{CW}}~=~-29.3(2)$~K. From the Curie constant one gets the effective copper magnetic moment as  $\mu_{\text{eff}}=\sqrt{3k_BC/N_A}$, where $k_B$ is the Boltzmann factor, and $N_A$ is Avogadro's number. We get $\mu_{\text{eff}}~=~1.920~\mu_B$ and 1.911~$\mu_B$ for  [001] and [100] directions, respectively. 
Here, $\mu_B~$ is the Bohr magneton. Using $g~=~\sqrt{S(S+1)}\mu_{\text{eff}}$ we get almost equal Lande $g$-factor values as $g~=~2.22$ and $g~=~2.20$ for  [001] and [100], respectively, which are common for Cu$^{2+}$ ions. 

Here we note, that in regular, collinear antiferromagnets the $\rchi$ vanishes in the direction where the local magnetic moments are aligned parallel or antiparallel to an external magnetic field and stays constant in the directions where the external field is perpendicular to the local magnetic moments~\cite{Singer1956}. Thus the local moments here may prefer the [001] direction, but the non-vanishing $\rchi$  could also just be a signature of the non-collinear order.

\subsection{$^{31}$P Knight shift}
\setlength{\parskip}{0em}
\begin{figure}[!]
	\includegraphics[width=0.40\textwidth]{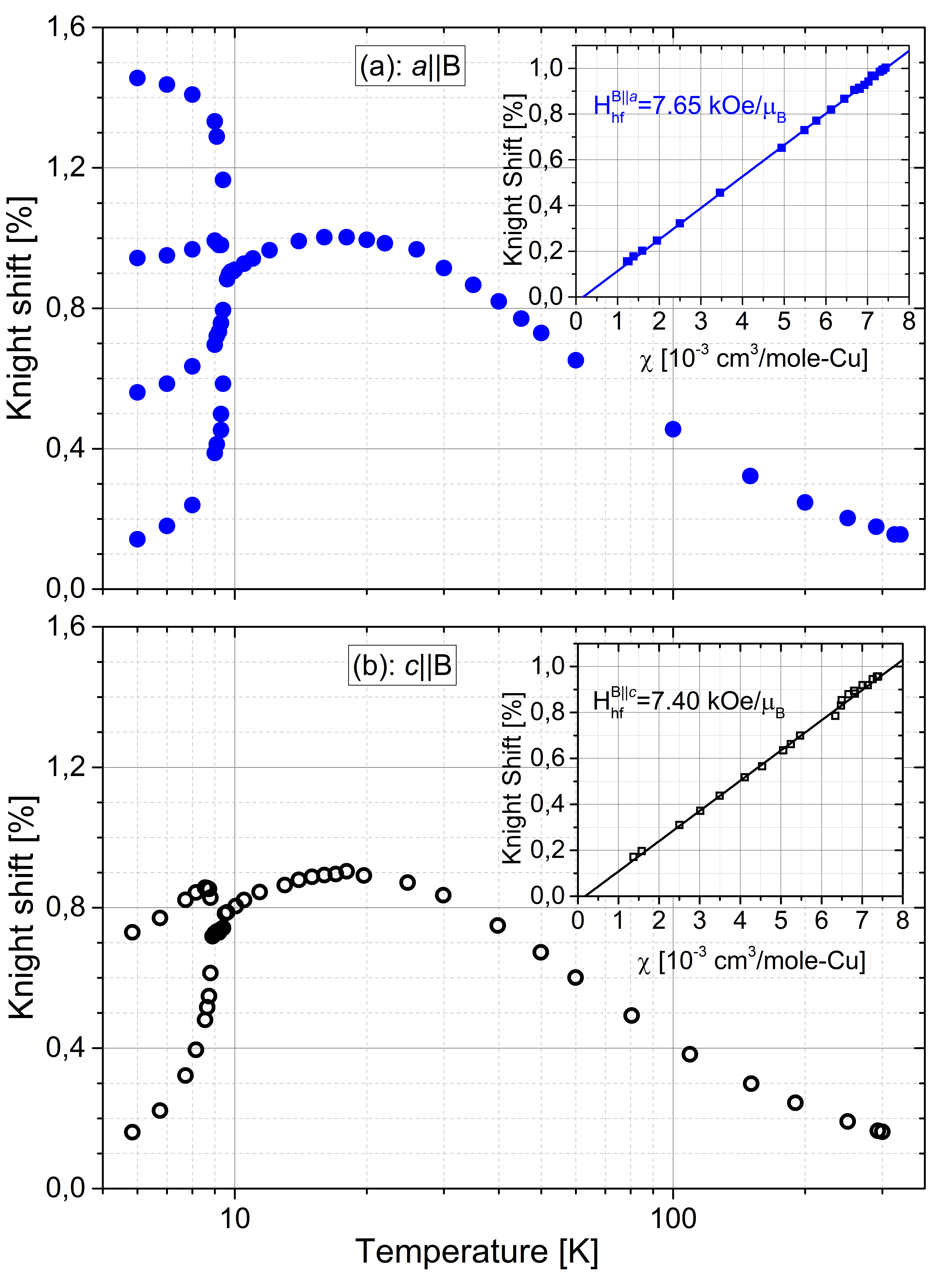} 
	\caption{\label{Knight}Temperature dependence of $^{31}$P Knight shift as measured when the single crystal is oriented with $a$ axis parallel to the external field $B$ (a, filled blue dots) and with $c$ axis parallel to $B$ (b, empty circles). The insets show the Clogston-Jaccarino plots giving quite close values to the hyperfine field in both the directions.}
\end{figure}

Temperature dependence of $^{31}$P Knight shift is presented in FIG.~\ref{Knight}.  Above the N\'{e}el temperature T$_N$~=~9.5~K, the Knight shift $K$(T) follows perfectly the magnetic susceptibility curve $\rchi$(T), as given~\cite{Clogston1961} by Eq.(\ref{C-J}):
\begin{equation} \label{C-J} 
K(T)=K_0+\frac{\textit{H}_{\text{hf}}}{N_A\mu_B}\rchi.
\end{equation}
\noindent
Here, $K_0$ is the temperature-independent shift, the chemical shift, and \textit{H}$_{\text{hf}}$ is the hyperfine coupling constant. 
From the $K $vs $\rchi$ plots (insets of  FIG.~\ref{Knight}) we found the hyperfine field values as \textit{H}$_{\text{hf}}$~=~7.65(5) kOe/$\mu_B$ for $a\parallel B$ direction and slightly smaller value,  \textit{H}$_{\text{hf}}$~=~7.40(5) kOe/$\mu_B$ for direction $c\parallel B$.
\begin{figure}[!]
		\includegraphics[width=0.40\textwidth]{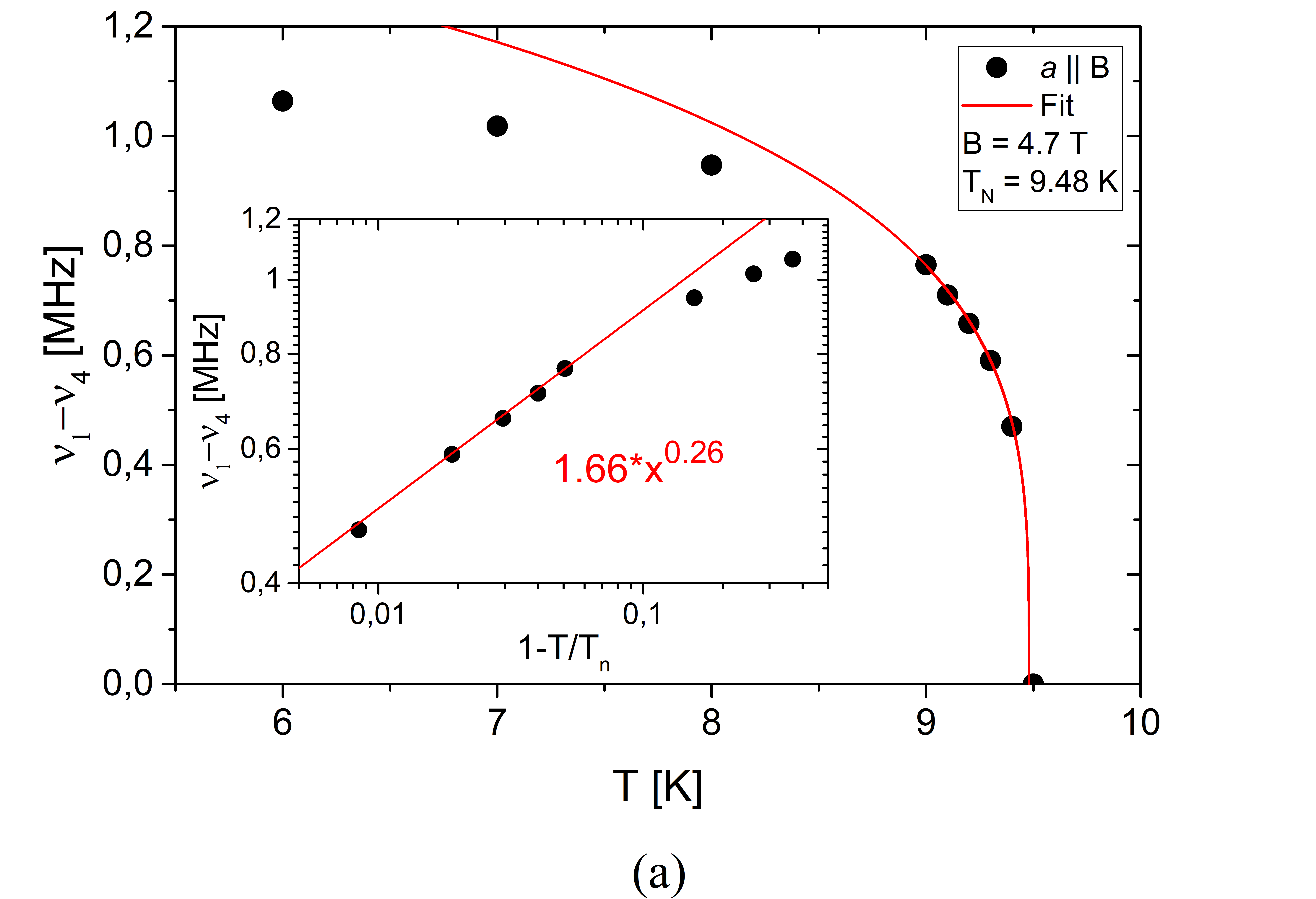} 
		\includegraphics[width=0.40\textwidth]{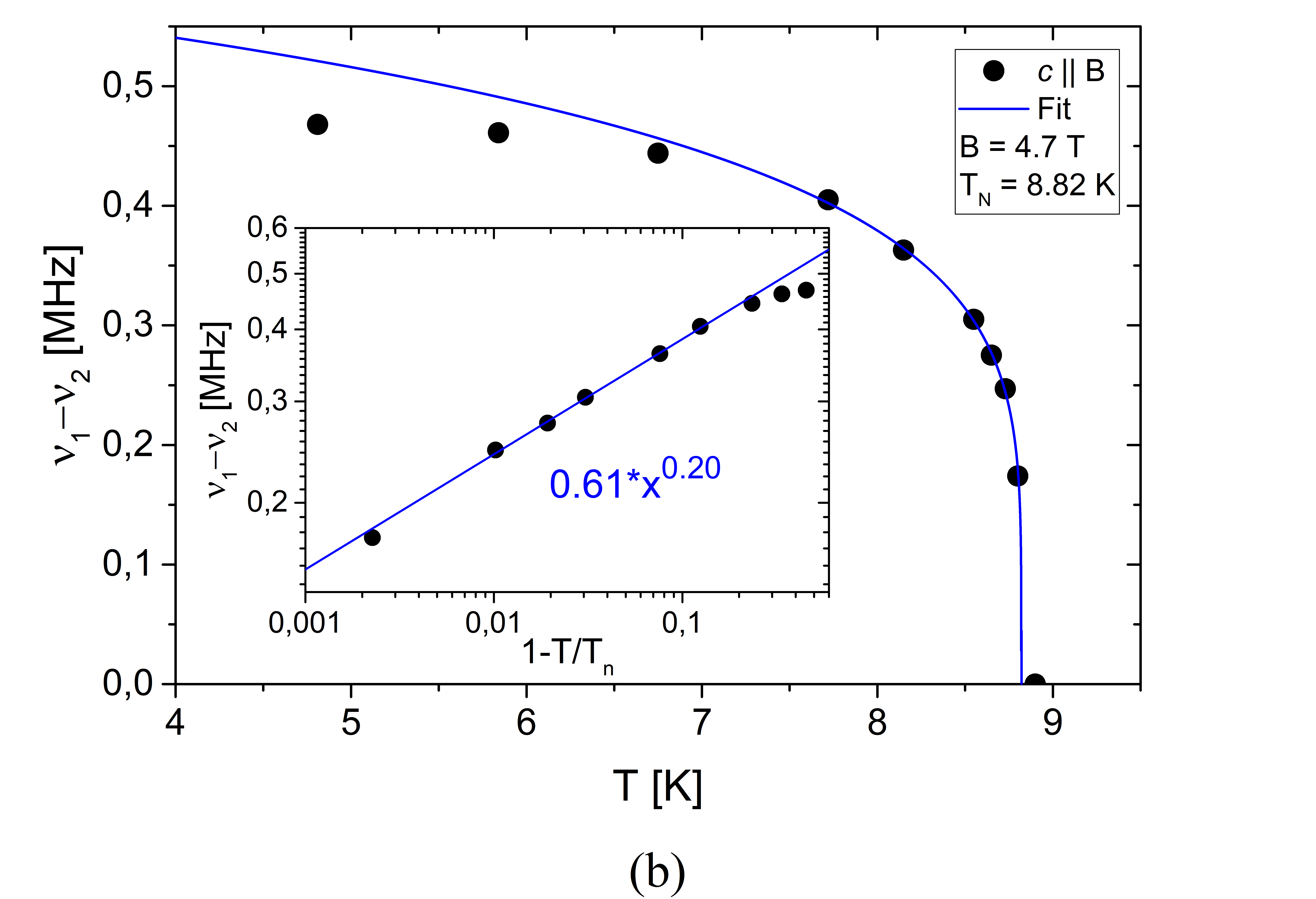} 
	\caption{\label{Tn}Temperature development of the splitting of the resonance lines below the ordering temperature in the orientation $a\parallel B$ (upper panel) and $a\parallel B$ (lower panel). The solid lines are fittings by Eq.~\ref{power} .}
\end{figure}

Below T$_N$, due to the  onset of local magnetic fields, the resonance line splits into four lines in the orientation $a\parallel B$ and into two lines when the crystal is oriented with $c\parallel B$.  
Following the approach given in Ref.~\cite{Islam2018} we will estimate the critical order parameter $\beta$ of the phase transition from the growth of the internal field $B_{int}$ in the vicinity of the ordering temperature. As a measure of the internal field we use the frequency difference between the most shifted resonance lines instead of the internal ones, which is proportional to the internal field by gyromagnetic ratio $^{31}\gamma/2\pi$~=~17.237~MHz/T. The temperature dependences of the frequency differences in the two orientations of the single crystal are given in FIG.~\ref{Tn}.

The temperature dependence of the frequency difference is fitted by the formula
\begin{equation}\label{power}
\Delta F(T)=\Delta F_0\left(1-\frac{T}{T_n}\right)^\beta.
\end{equation} 

The best fit was obtained with  T$_N$~=~9.48~K, $\beta$~=~0.26 for the orientation $a\parallel B$,  and  T$_N$~=~8.82~K, $\beta$~=~0.20 for  $c\parallel B$.  The critical exponent $\beta$ values, if compared to some theoretical values (as given \textit{e. g.} in Ref.~\cite{Nath2009}), indicate that the ordering scheme is the closest to the 2D XY case. 
\subsection{$^{31}$P NMR of powder sample} 
\begin{figure}[!]
	\includegraphics[width=0.40\textwidth]{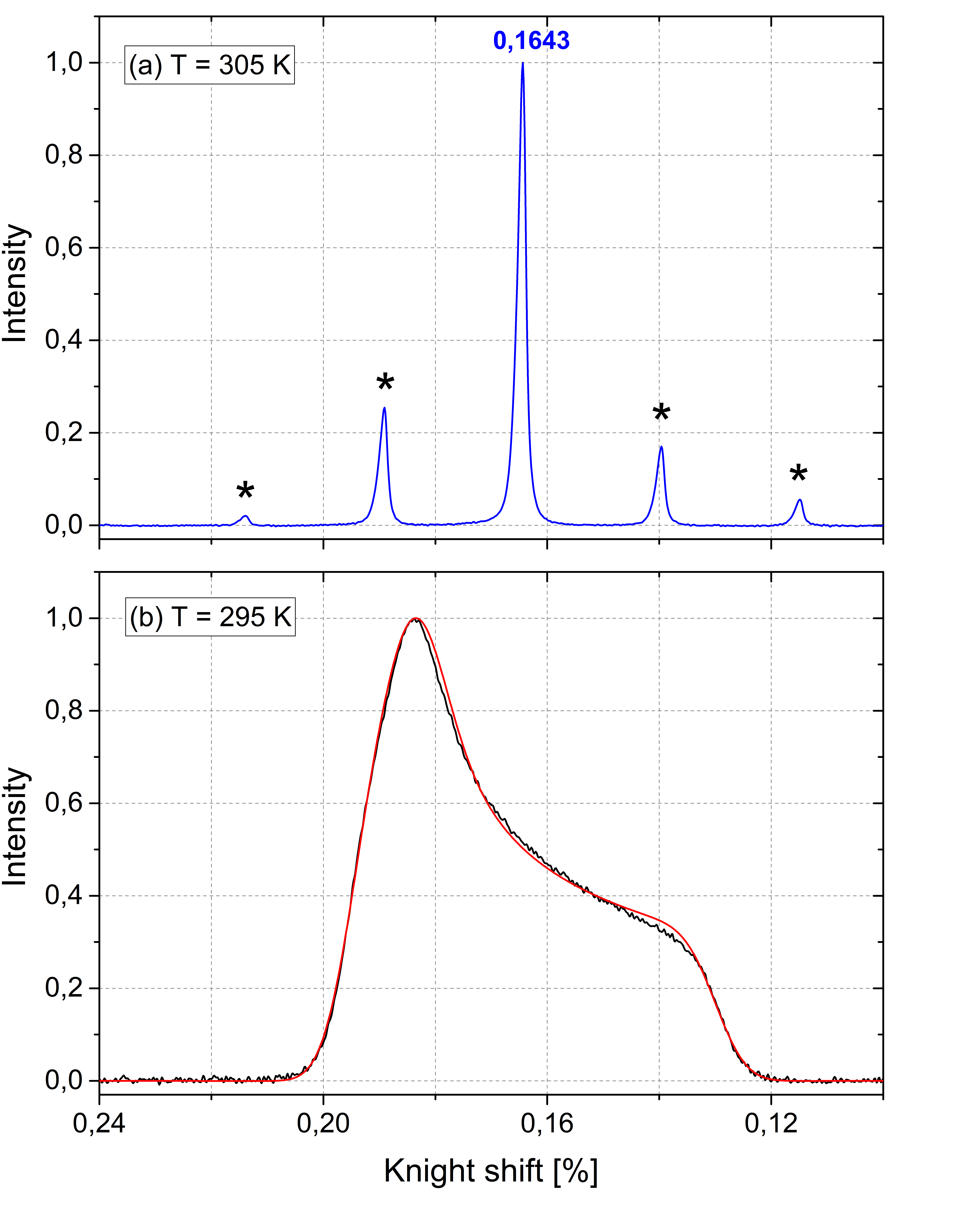} 
	\caption{\label{powder}(a): $^{31}$P MAS NMR of BaTCPO powder sample at T~=~305~K with a spinning frequency $\nu$~=~25~kHz. 
The NMR spectrum of rotating sample consists of a center band at isotropic Knight shift K$_{\text{iso}}$~=~0.1643~\%, and of a number of spinning sidebands denoted by asterisks at multiples of the spinning frequency from the center band. (b): $^{31}$P NMR of a static powder sample at slightly lower temperature T~=~295~K. The red line is a fit to the powder line with the eigenvalues of nearly axially symmetric Knight shift tensor: K$_{11}~=~0.1952$~\%, K$_{22}~=~0.1832$~\%, K$_{33}~=~0.1298$~\%.}
\end{figure}

Before starting $^{31}$P analysis of single crystal we recorded the spectrum of a powder sample.
FIG.~\ref{powder}(a) shows $^{31}$P NMR spectrum recorded with magic angle spinning~\cite{MAS-NMR_reference} of the sample. The spectrum shows a single sharp line at isotropic magnetic shift. The spectrum of a static sample (FIG.~\ref{powder}(b)) shows a typical powder line shape with singularities at the principal values of the Knight shift tensor: 
\begin{equation}\label{PAS}
	K_{PAS}=
	\begin{pmatrix} 1+K_{11} & 0 & 0 \\ 0 & 1+K_{22} & 0 \\ 0 & 0 & 1+K_{33} \end{pmatrix},
\end{equation}
with K$_{11}=0.1952$ \%, K$_{22}=0.1832$ \%, K$_{33}=0.1298$ \%.  These values give us  the reference for the interpretation of the rotation patterns of the single crystal. 

\subsection{Orientation of the $^{31}$P Knight shift tensor} 

Determination of tensor orientation in a single crystal is not very easy task~\cite{Mehring1976}. For that, in general case, one needs to record resonance frequencies rotating the sample around three different axes. Depending on symmetry or having some principal values of the shift tensor pre-determined, the number of necessary rotation patterns may be smaller. After obtaining those rotation patterns, one needs to find unitary transformation that will transform the principal axis system (PAS) into the crystal frame. 
As usual, the Hamiltonian of spin-$\frac{1}{2}$ nucleus consists of the Zeeman and the Knight shift interaction
\begin{equation} \label{H}
\mathcal{H}=\mathcal{H}_Z+\mathcal{H}_K=\gamma \textbf{H}_0\cdot \textbf{K} \cdot \textbf{I},
\end{equation}
where $\textbf{I}=(I_x,I_y,I_z)^T$, $\textbf{H}_0 = (0,0,H_0)$, and $I_x,I_y,I_z$ are Pauli matrices:
\begin{equation}
I_x=\frac{\hbar}{2}\begin{pmatrix} 0 & 1 \\ 1 & 0 \end{pmatrix}; 
I_y=\frac{\hbar}{2}\begin{pmatrix} 0 & i \\ -i & 0 \end{pmatrix}; 
I_z=\frac{\hbar}{2}\begin{pmatrix} 1 & 0 \\ 0 & -1 \end{pmatrix}.\\
\end{equation}
Three successive rotations, each characterized by three Euler angles, 
transform the Hamiltonian from the principal axes frame (PAS) into the laboratory frame:
\begin{eqnarray*} \label{H}
 \begin{array}{c}K_{\scriptstyle PAS}\\ \scriptstyle(1,2,3)\end{array}\underrightarrow{ \scriptstyle R_p(a_1,a_2,a_3)}\begin{array}{c}K^*\\ \scriptstyle(a,b,c)\end{array}\scriptstyle \underrightarrow{R_i(b_1,b_2,b_3)}\begin{array}{c} K_g\\ \scriptstyle (x_g,y_g,z_g) \end{array}\scriptstyle \underrightarrow{R_g(c_1,c_2,c_3)}\begin{array}{c} K\\ \scriptstyle(x,y,z)\end{array} 
\end{eqnarray*}
 There are four frames of reference: PAS with diagonal tensor ($ K_{PAS} $), crystal frame ($ K^* $), goniometer frame ($ K_g $), and laboratory frame (K). The transformations between them are $R_p$, $R_i$, $R_g$ with the corresponding Euler angles. We used the conventional Euler ZYZ rotation of the Hamiltonian~\cite{easyspin}.

\begin{figure*}[!]
	\begin{center}
		\includegraphics[width=0.75\textwidth]{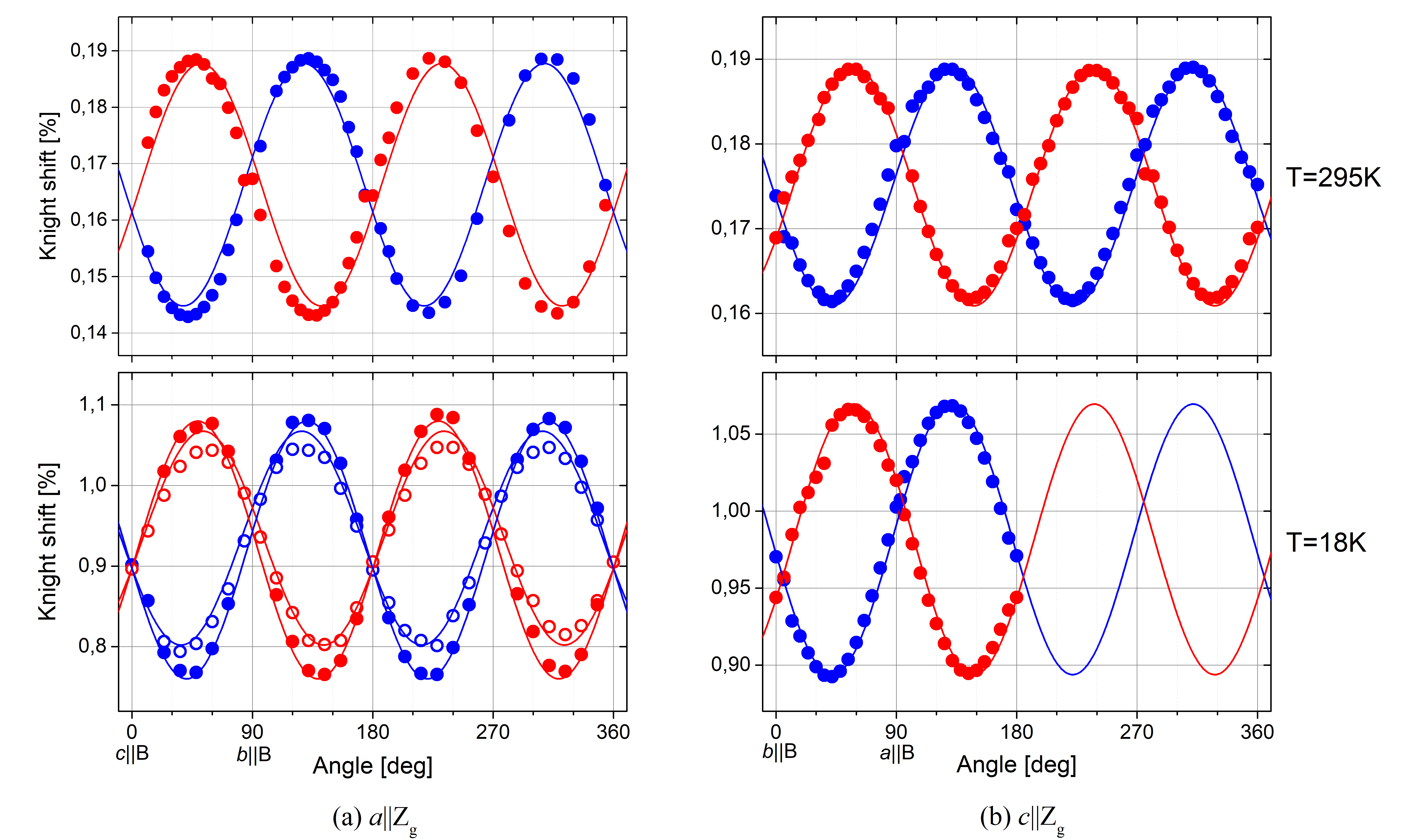} 
		\caption{\label{rotation} $^{31}$P NMR resonance frequencies by rotating BaTCPO single crystal around the $a$ (a) and $c$ axis (b) at T=295~K (upper panels) and T=18~K (lower panels). The full lines correspond to the angle dependencies according to the parameters given in Table~\ref{details}. Due to symmetry, eight different sites in the unit cell contribute only to two different rotation patterns. The red lines in panel (a) correspond to the sites 1, 3, 6 and 7 (see Table~\ref{details}), and to the sites 2, 3, 6 and 8 in panel (b), the rest of the sites are given by blue lines.}
		\vspace{5mm}
	\end{center}
\end{figure*}

\begin{figure}[ht!]
		\includegraphics[width=0.4\textwidth]{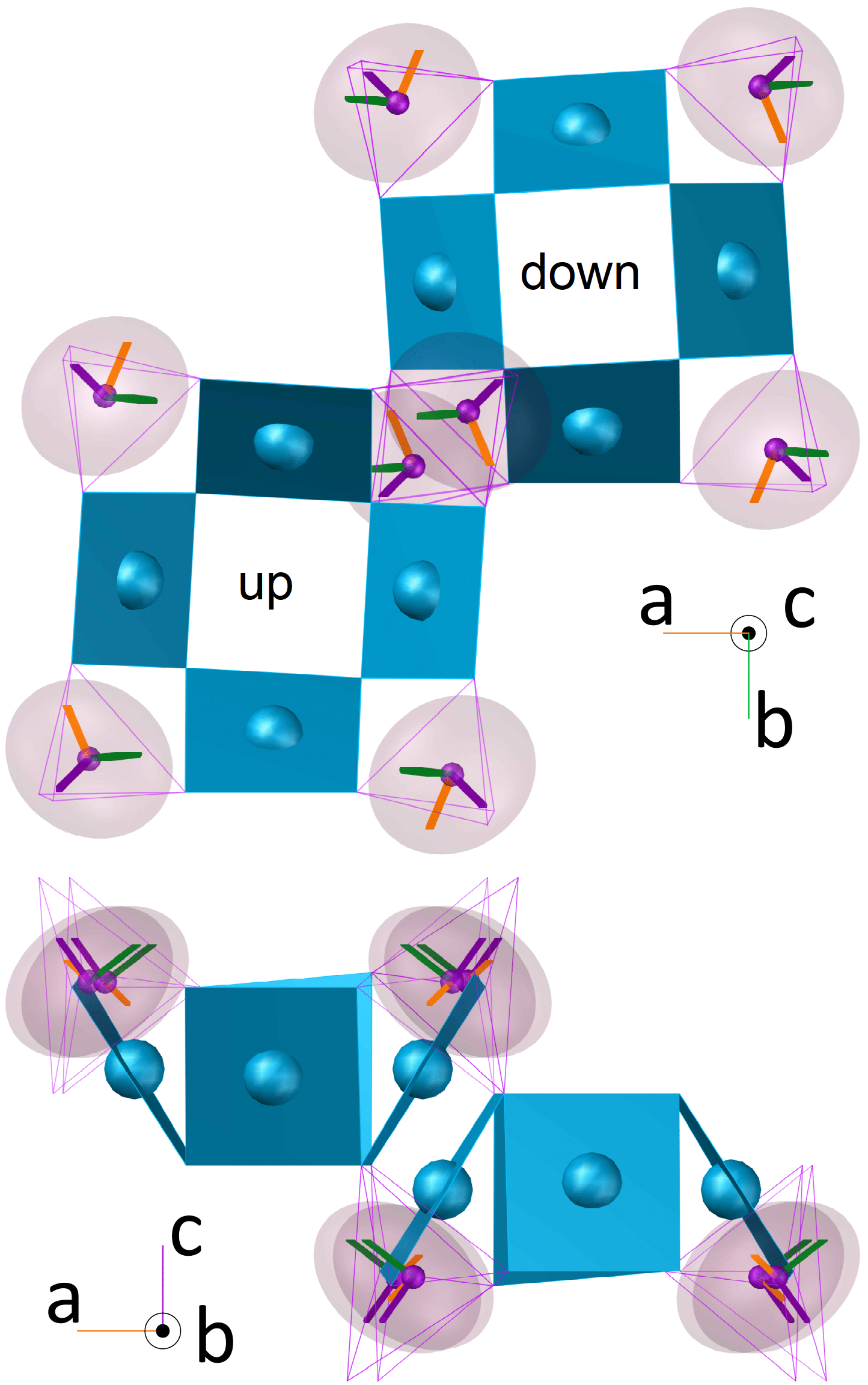} 
		\caption{\label{BTCP_295K} Orientations of the Knight shift tensor at eight phosphorus locations of the unit cell
at temperature T~=~295~K. The tensors have three principal vectors - green, orange, dark-violet, which correspond to the $K_{PAS}$ values $K_{11}$, $K_{22}$, $K_{33}$, respectively.}
\end{figure}

In BaTCPO there are eight different positions of the phosphorus ions in the unit cell, each giving the resonance corresponding to different transformation $R_p$. 
The rotation patterns around the $c$ and $a$ axis at temperatures T~=~295~K and T~=~18~K are shown in FIG.~\ref{rotation}. 
The angles $b_1$, $b_2 $, $b_3$ and $c_1$, $c_2 $, $c_3$ are unique for each experiment; $a_1$, $ a_2 $, $a_3$ are desired eight sets of the Euler angles transforming the tensor in PAS to the crystal frame for each phosporus site. It turns out that the principal axis $K_{33}$ of the Knight shift tensor is tilted by 45 degrees from the crystal $c$ axis. 
Schematics of the Knight tensor's eight orientations are presented in Fig.~\ref{BTCP_295K}.

\begin{table*}
	\caption{\label{tab:18K}(a) and (b): the Euler angles for transforming BaTCPO single crystal shift tensors from crystal frame to the laboratory frame. Tables (c) and (d) show the Euler angles for transforming the tensors from PAS to the crystal frame. Tensor components of PAS ($ K_{11} $,$ K_{22} $,$ K_{33} $) are in $\%$ units of $\omega_L$.}
	\label{details}
	\begin{ruledtabular}
	\begin{tabular}{|c|c c c|c c c|c|c c c|c c c|c c c|}
		\multicolumn{7}{|c|}{(a) T=295K} & T & & (c) 295K & & & (d) 18K & & \multicolumn{3}{c|}{T=295K} \\
		\hline
		& $c_1$ & $c_2 $ & $c_3$ &$b_1$ & $b_2 $ & $b_3$ &nr & $a_1$ & $ a_2 $ & $a_3$ & $a_1$ & $a_2$ & $a_3$  & $K_{11}$ & $K_{22}$ & $K_{33}$  \\ 
		\hline 
		c$\parallel$z$_g$ & (-12:372) & 90 & 0 & 85 & 0 & 0 &1 &  30 &  45 & -45 & 25 & 45 &-40 & $0.198$ & $0.185$ & $0.128$ \\
		a$\parallel$z$_g$ & (-12:372) & 90 & 0 & 0 & 90 & 0 &2 & -30 &  45 &  45 &-25 & 45 & 50 &   &   &  \\
		&           &    &   &   &    &   &3 &  30 &  45 & 135 & 25 & 45 & 140&   &   &  \\
		\cline{1-7} \cline{15-17} 
		\multicolumn{7}{|c|}{(b) T=18K} & 4 & -30 &  45 & -135  & -25 & 45 & -130 & \multicolumn{3}{c|}{T=18K}\\
		\cline{1-7} \cline{15-17} 
		& $c_1$ & $c_2 $ & $c_3$ &$b_1$ & $b_2 $ & $b_3$ & 5 & 30 & 135 &  45 &  25 & 135 & 50 &$K_{11}$ & $K_{22}$ & $K_{33}$ \\ 
		\cline{1-7} \cline{15-17}
		c$\parallel$z$_g$ & (-12:372) & 90 & 0 & 80 & 0 & 0 & 6 & -30& 135 & -45 & -25 & 135 & -40 & $1.14$ & $1.05$ & $0.67$ \\
		a$\parallel$z$_g$ & (-12:372) & 90 & 0 & 0 & 90 & 0 & 7 & 30 & 135 &-135 & -25 & 135 & 140   &   &   & \\ 
		&           &    &   &   &    &   & 8 & -30& 135 & 135 &  25 & 135 & -130  &   &   & \\
	\end{tabular}
	\end{ruledtabular}
\end{table*}

\subsection{Spin-lattice relaxation results}

\begin{figure}[!ht]
		\includegraphics[width=0.5\textwidth]{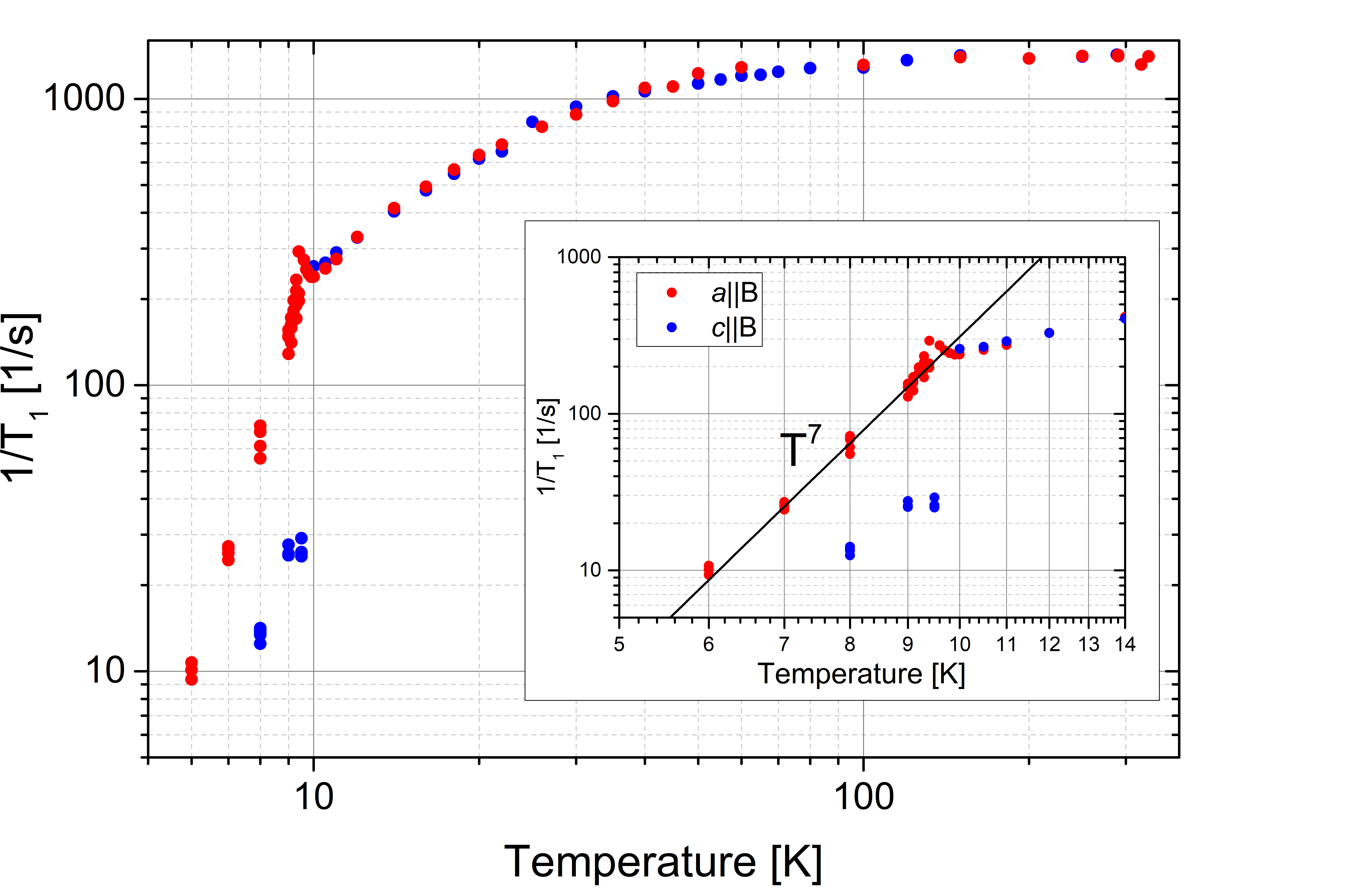} 
		\caption{\label{relax} Temperature dependence of $^{31}$P spin-lattice relaxation rate in directions [100] (red dots) and [001] (blue dots). The inset shows that relaxation rate 1/T$_1$ is proportional to T$^7$ below T$_N$.}
\end{figure}

Spin-lattice relaxation $T_1$was measured with inversion-recovery pulse sequence at magnetic fields along [001] and [100] directions (Fig.~\ref{relax}). The magnetization recovery was exponential throughout all the measurements:
\begin{equation}
    M(\tau)=M_0(1-A\exp(-\tau/T_1)),
\end{equation}
where $M(\tau)$ is the magnetization at delay $\tau$ after inversion, $M_0$ is the equilibrium magnetization, and $A~\le$~2 is a constant depending on the accuracy of the inversion.

The relaxation rate at T $>$ 60 K is almost constant which is typical for a paramagnetic material, where the relaxation is caused by the fluctuation of the magnetic moments. Before the phase transition at T$_N$, a sharp spike occurs in the relaxation speed which is connected to the rapid slowing of the fluctuations. Below T$_N$, relaxation speed decreases sharply proportional to T$^7$. In case of $c~\parallel~B$, the relaxation rate seems to have a discontinuity in close vicinity of T$_N$. Here we note, that 1/T$_1$ $\propto$ T$^7$ was also observed for SrTCPO~\cite{Islam2018}.

In paramagnetic temperature region we can use the Moriya's theory of relaxation \cite{Moriya}, where:
\begin{equation}
	\frac{1}{T_1}=\frac{\gamma ^2_N \sqrt{2\pi} S(S+1)}{3\omega_E z'} H^2_{hf},
\end{equation}

Here $\gamma_N$ is the nuclear gyromagnetic ratio, $ S $ is the nuclear spin, $H_{hf}$ is the hyperfine field as in Eq.~\ref{C-J}, $z'=2$ is the number of nearest Cu$^{2+}$ neighbors for the nucleus and
$\omega_E=(|J| k_B/ \hbar) \sqrt{2 z S(S+1)/3}$ is the Heisenberg exchange frequency (in units rad$^{-1}$), where $z$~=~2  is the number of Cu$^{2+}$ ions as nearest neighbors to $^{31}$P, $S$~=~1/2 is the electronic spin, and $J$ is the exchange interaction.

With the hyperfine field value of $H_{hf}$~=~7.650 kOe/$\mu_B$ and the relaxation rate for the paramagnetic region 1/T$_1$ = 1410 s$^{-1}$ we get an estimate for the exchange interaction inside the square cupola $J/k_B$~=~35~K with an exchange frequency of $\omega_E$~=~4.5$\cdot10^{12}$~rad/s. This value is in coherence with previous results of $J$~=~3.0~meV~=~34.8~K~\cite{Kimura2016magneto}.

\subsection{Local magnetic structure in the ordered state}

\begin{figure}[!]
		\includegraphics[width=0.45\textwidth]{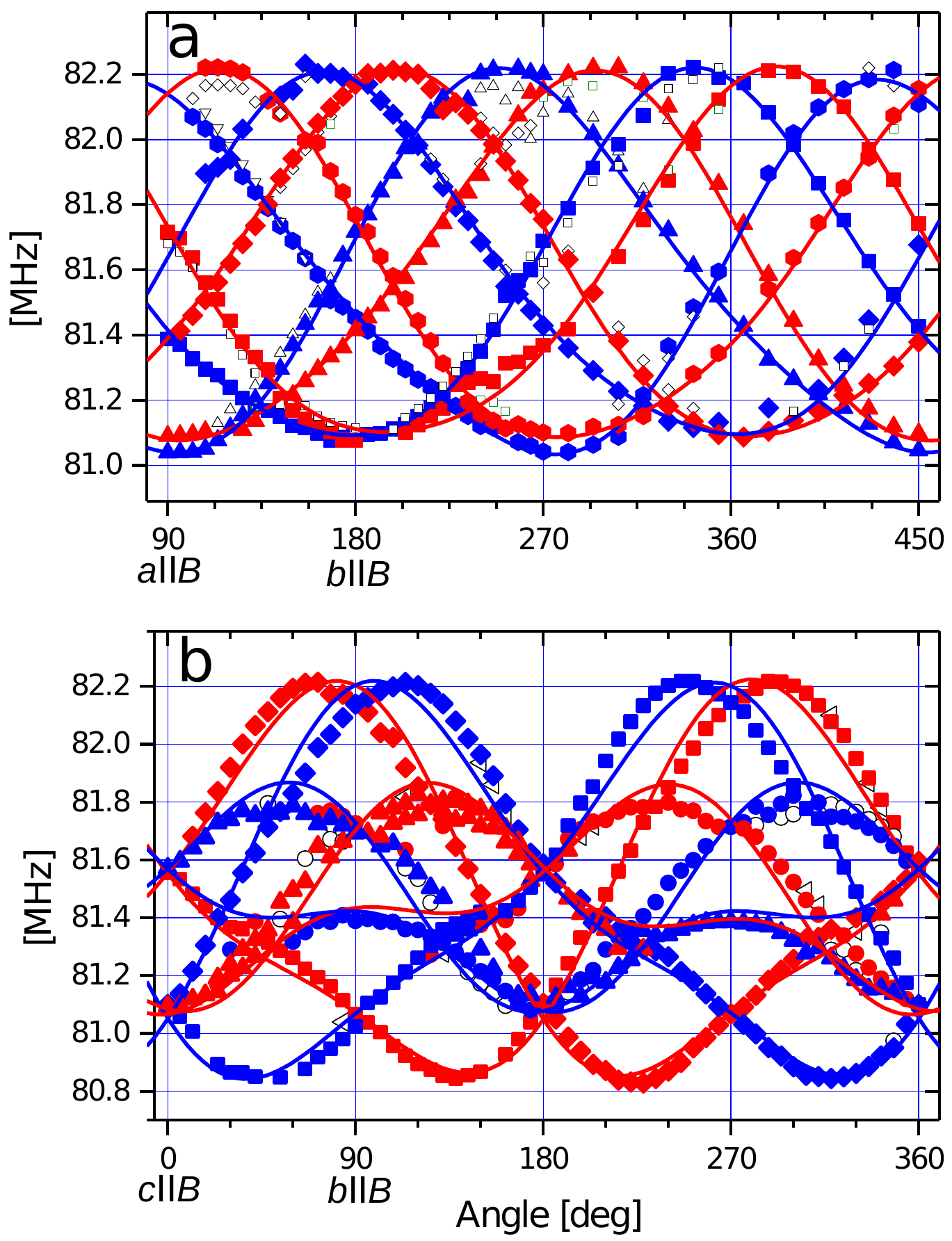} 
	\caption{\label{rot_6K}Rotation patterns of $^{31}$P NMR frequencies of BaTCPO single crystal at temperature T~=~6~K rotating 	around [001] (a) and [100] (b). The lines are approximations with Eq.~\ref{lahendus} using data in Table~\ref{tab:6K}. 
	The main directions of a crystal $a\parallel$B, $b\parallel$B, and $c\parallel$B, $b\parallel$B are noted in the respective $x$ axes. The blue symbols are assigned to the sites in ``up'' cupola, the red symbols to the sites in ``down'' cupola.} 
\end{figure}

The resonance frequencies by rotation of the single crystal around the [001] and [100] at T~=~6~K are given in Fig.~\ref{rot_6K}. The results are in coherence with the Knight shift temperature dependence (FIG.~\ref{Knight}). Once the magnetic field is turned around the $c$ axis, one can see that when $a\parallel$~B, there are four different magnetic field projections in the AFM region as given in FIG.~\ref{Knight}(a). When the crystal is oriented $c\parallel$~B, there are two different local field projections to the external field direction as given in FIG.~\ref{Knight}(b). Rotating the sample around $c$ and $a$ gives eight different rotation patterns for eight phosphorus ions in the unit cell. Each rotation pattern can be described by the equation:
\begin{equation} \label{lahendus}
F=K+L\cos(\alpha-\alpha_1)+M\cos(2(\alpha-\alpha_2)),
\end{equation}
where the constant term $K$ is the Larmor frequency plus average chemical shift; the second term describes the angle dependence of the local field projection to the external field direction. The third term describes the angle dependence of the resonance frequency due to turning of 
the chemical shift tensor. 

The rotation patterns around the [001] axis show two sets of lines (blue and red). The phase shift within the set is 90 degrees. The red lines are shifted from $b\parallel$~B by +16~degrees, while the blue lines are shifted -16~degrees. We can assign the blue lines to the phosphorus ions in ``up'' cupola and the red lines to the ions in ``down'' cupola. The rotation patterns of the crystal around the [100] (FIG.~\ref{rot_6K}(b), Table~\ref{tab:6K}b) are not so well resolved. Here, the approximation of the frequencies by Eq.~\ref{lahendus}  is not particularly good. A possible reason might be that the magnetic structure in 4.7~T magnetic field at $c\parallel$~B direction is not yet well settled at temperature 6~K. We found above (see FIG.~\ref{Tn}) that the ordering temperature in the direction $c\parallel$~B was T$_N$~=~8.8~K, while in case of $b\parallel$~B we had T$_N$~=~9.5~K. 

Despite of that, one can clearly see the rotation patterns with two different amplitudes as expected for two different local field projections along the $b$ axis. The assignment of the resonances to ``up''  and ``down'' cupola is not unique. For example, we cannot distinguish the cases, where all the local field directions of one cupola have positive projection to the $c$ axis and the moments of the other cupola have negative projection, from the case, where the ions of one cupola have two positive and two negative projections and the field on ions of the other cupola have, respectively, two negative and two positive projections to the $c$ axis. The assignment in FIG.~\ref{rot_6K}(b) corresponds to the local field configuration as given below in FIG.~\ref{str:6K}.

\begin{table}[!]
	\caption{\label{tab:6K}The fitting parameters $K$, $L$, $M$, $\alpha_1$, and $\alpha_2$ of the rotation patterns according to Eq.~\ref{lahendus} 
The upper table corresponds to the rotation of the crystal around the $c$ axis (as shown on FIG.~\ref{rot_6K}(a)), lower table corresponds to rotation  around the $a$ axis (FIG.~\ref{rot_6K}(b)) at temperature T~=~6~K.}
	\begin{ruledtabular}
		\begin{tabular}{c  c c c c c} 
			nr & $K$ & $L$ &$\alpha_1$ & $M$ & $\alpha_2$ \\ [0.5ex] 
			\hline
			1 & 81.59 & 0.56 & 347 & 0.075 & 155\\ 
			2 & 81.59 & 0.56 & 20 & 0.075 & 25\\ 
			3 & 81.58 & 0.56 & 80 & 0.075 & 50\\ 
			4 & 81.59 & 0.56 & 110 & 0.075 & 125\\ 
			5 & 81.59 & 0.55 & 170 & 0.075 & 155\\ 
			6 & 81.595 & 0.56 & 198 & 0.065 & 30\\ 
			7 & 81.59 & 0.575 & 258 & 0.078 & 50\\ 
			8 & 81.60 & 0.56 & 288 & 0.065 & 130\\  
		\end{tabular}
		\begin{tabular}{c  c c c c c} 
			nr & $K$ & $L$ &$\alpha_1$ & $M$ & $\alpha_2$ \\ [0.5ex] 
			\hline
			1 & 81.47 & 0.62 & 294 & 0.17 & 87\\ 
			2 & 81.47 & 0.62 & 246 & 0.17 & 97\\ 
			3 & 81.45 & 0.29 & 212 & 0.17 & 72\\ 
			4 & 81.44 & 0.29 & 148 & 0.17 & 115\\ 
			5 & 81.47 & 0.62 & 114 & 0.17 & 84\\ 
			6 & 81.47 & 0.62 & 66 & 0.17 & 95\\ 
			7 & 81.45 & 0.29 & 32 & 0.17& 69\\ 
			8 & 81.45 & 0.29& 328 & 0.17 & 111\\
		\end{tabular}
	\end{ruledtabular}
\end{table}

\begin{figure}[!]
		\includegraphics[width=0.4\textwidth]{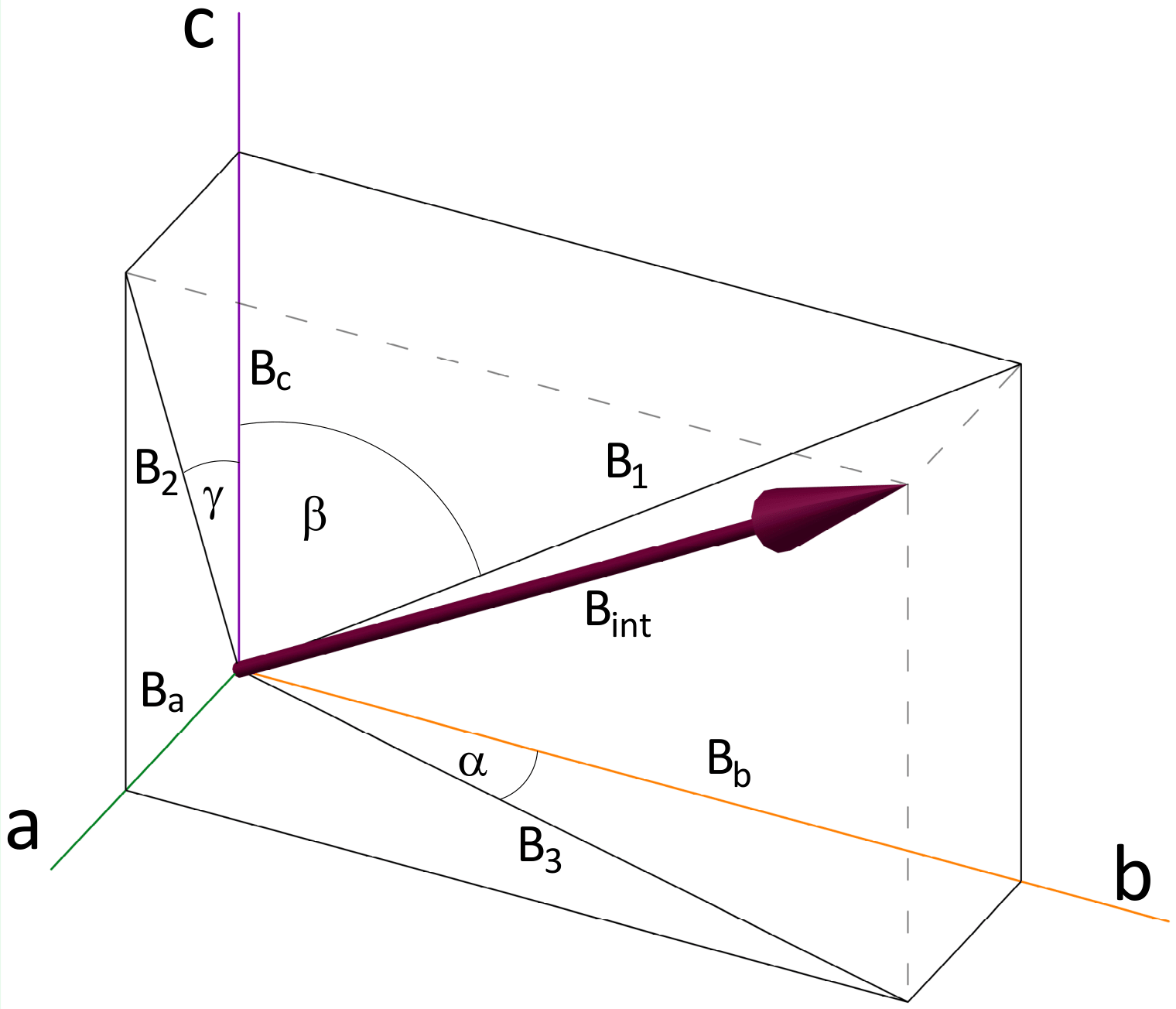} 
		\caption{ \label{axes}Scheme of local field direction on a phosphorus ion in BaTCPO crystal. B$_\text{int}$ is the local field, 
		B$_a$, B$_b$, and B$_c$ are the projections of B$_\text{int}$ to the crystal $a$, $b$, and $c$ axis, respectively. B$_1$, B$_2$, and B$_3$ are the projections of B$_\text{int}$ to the $bc$, $ac$, and $bc$ planes, respectively. The latter amplitudes can be found from the rotation pattern parameters given in Table~\ref{tab:6K}. }
\end{figure}

The analysis of the data given in Table~\ref{tab:6K} can be carried out using the scheme of the local field direction as given in FIG.~\ref{axes}. Three cosine amplitudes $L$ in Table~\ref{tab:6K} correspond to the local field projections B$_1$, B$_2$, and B$_3$ in FIG.~\ref{axes}.
Using the gyromagnetic ratio of $^{31}$P $\gamma/2\pi$~=~17.237~MHz/T, we find B$_1$~=~36~mT, B$_2$~=~32.5~mT, and B$_3$~=~16.8~mT. The angles $\alpha$~=~$\pm$~16, $\beta$~=~$\pm$~66, and $\gamma$~=~$\pm$~32 degrees. It is not difficult to calculate the projections B$_a$~=~8.9~mT, B$_b$~=~31.2~mT,  and B$_c$~=~14.6~mT, as well as the module of the internal field B$_\text{int}$~=~35.6~mT.

As noted above, unique assignment of the resonances to certain phosphorus in the unit cell is not possible. One possible local field configuration is given in FIG.~\ref{str:6K}. 
 
\begin{figure}[!]
	\includegraphics[width=0.4\textwidth]{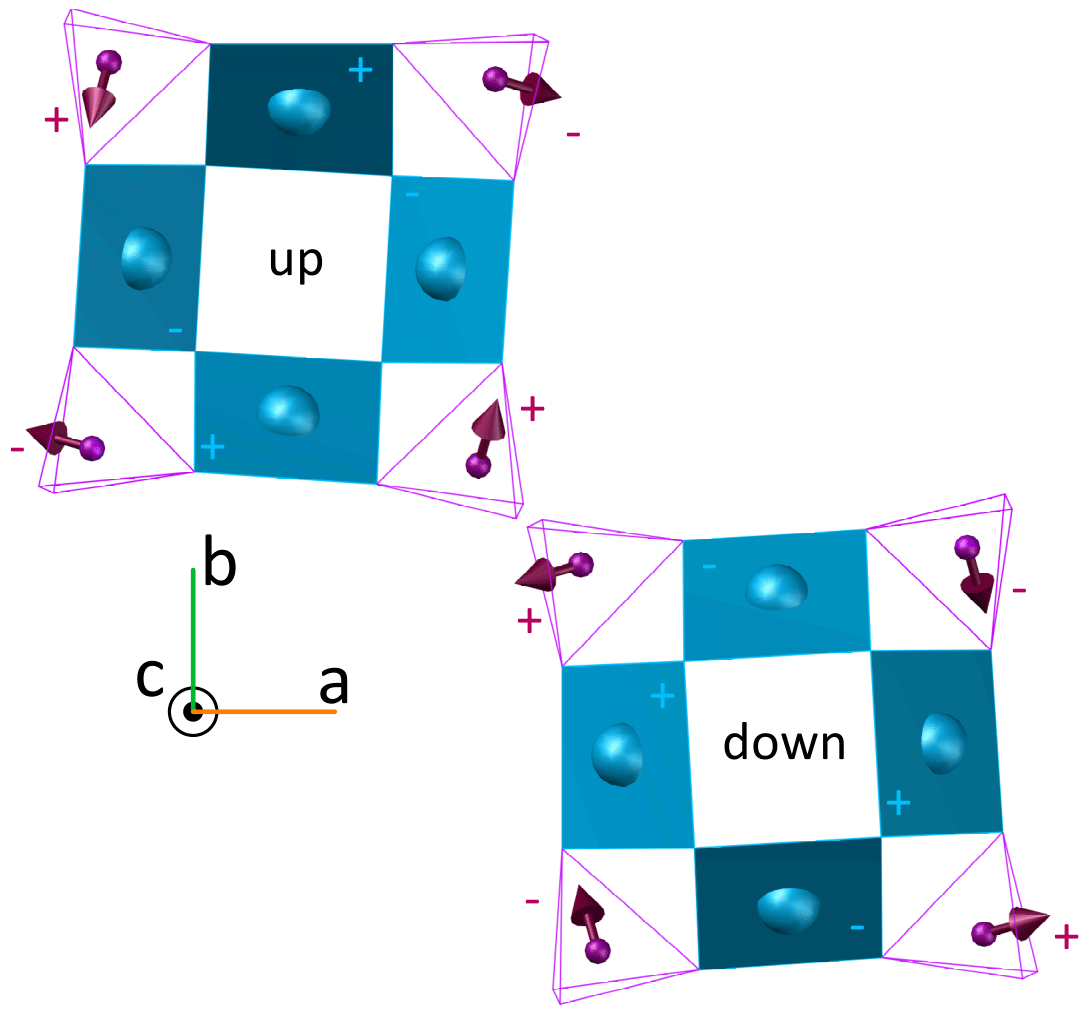}
	\caption{\label{str:6K}Possible configuration of induced static magnetic fields at phosphorus ions in BaTCPO unit cell. Red arrows represent the directions and size of the static magnetic fields, plus and minus sign mean the field direction to the front or to the back of the figure plane. Blue squares are Cu$^{2+}$ ions and the purple tetrahedrons show P ions.}
\end{figure}

The internal field at phosphorus ions consists of two components:
\begin{equation} \label{Bint}
B_\text{int}=B_\text{hf}+B_\text{dip},
\end{equation}
where $B_\text{hf}$ is transferred hyperfine field, and $B_\text{dip}$ is the dipolar field from the magnetic moments of Cu$^{2+}$ ions. In the following we assume that $B_\text{hf}$  is very well cancelled in AF ordered state, and the local magnetic field at phosphorus is due to dipolar field  of Cu$^{2+}$ ions.

We did calculate the dipolar field at each phosphorus ion of the unit cell assuming the \textit{two} magnetic structures proposed in Refs.~\cite{Kimura2016magneto,Babkevich2017}. In the calculation we did sum the dipolar field from every Cu$^{2+}$ inside a sphere of 50~\AA  \ around given phosphorus. At that we took into account that the unit cell of magnetic structure is doubled along $c$ direction, \textit{i. e.} the magnetic moments of every other layer along the $c$ axis were reversed. As a result, we obtain the following dipolar field projections along the $a$, $b$, and $c$ axis 
\begin{align*}
	\text{a) }&B_a=37.5~\text{mT; }B_b=90.1~\text{mT; }B_c=11.2~\text{mT, }\\
	&B_\text{int}=97.7~\text{mT;}
\end{align*}
	for the structure noted $\varGamma_3(1)$, and 
\begin{align*}
	\text{b) }&B_a=42.2~\text{mT; }B_b=36.0~\text{ mT; }B_c=0.75~\text{ mT, }\\
	&B_\text{int}=55.5~\text{ mT;}
\end{align*}
 for the structure noted $\varGamma_3(2)$.
 
 In comparison, our experiments above give the values: 
\begin{align*}
&B_a=8.9~\text{mT; }B_b=31.2~\text{mT; }B_c=14.6~\text{mT; } \\
&B_\text{int}=35.6~\text{mT.} 
\end{align*}
\begin{figure}[!]
		\includegraphics[width=0.375\textwidth]{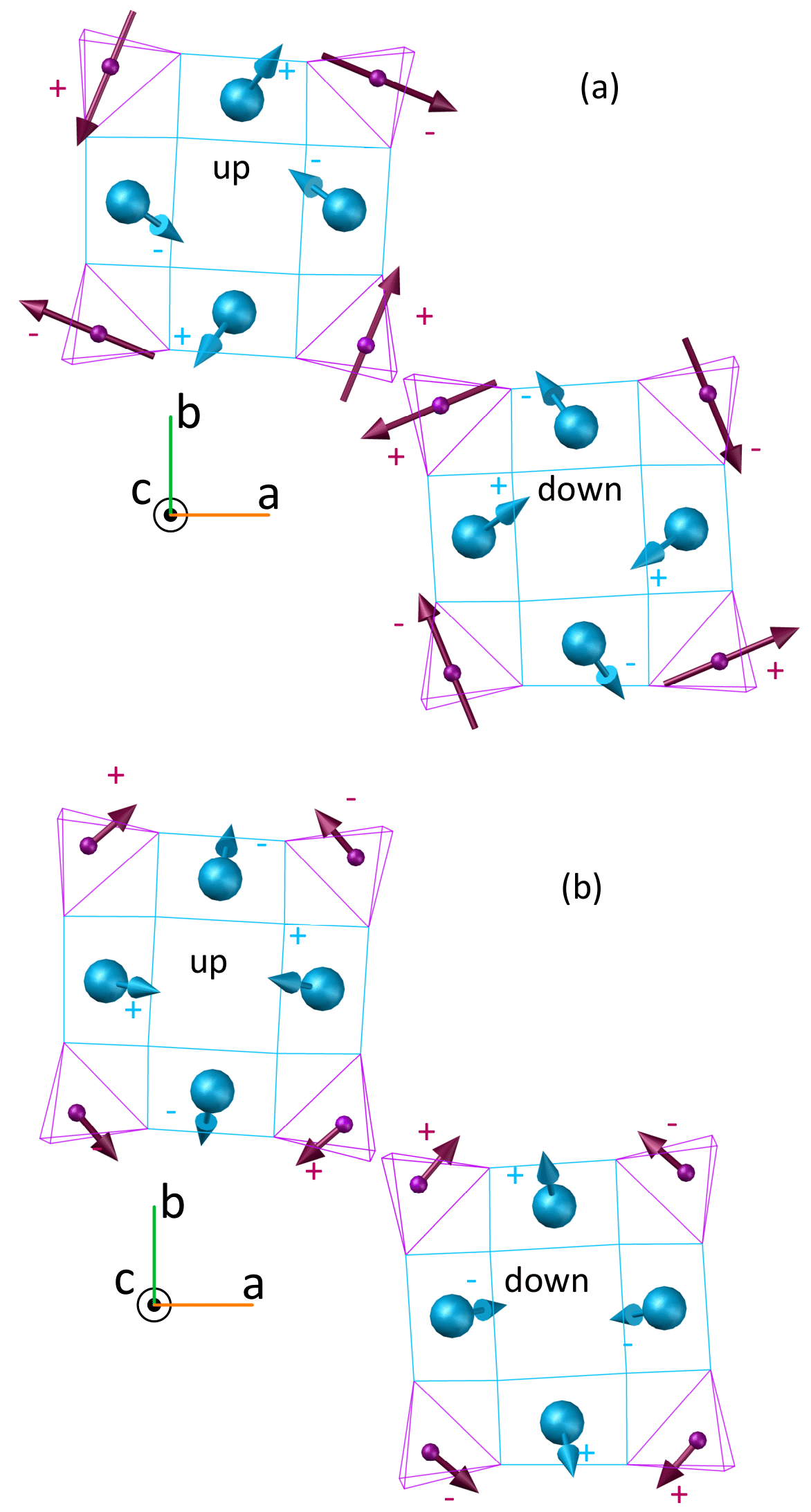} 
		\caption{ \label{dipolar}Calculated dipolar field direction at phosphorus ions of the BaTCPO unit cell (a): assuming  structure of Cu$^{2+}$  magnetic moments $\varGamma_3(1)$ (given in Ref.~\cite{Kimura2016magneto} ), and (b): assuming the structure $\varGamma_3(2)$. The red arrows represent the dipole fields. The length of the arrows corresponds to the relative value of dipolar field in both cases and that in FIG.~\ref{str:6K}. Blue arrows represent directions of the Cu$^{2+}$ magnetic moments and their sizes in the two cases.}
\end{figure}

Calculated dipolar field directions at the phosphorus ions are given in FIG.~\ref{dipolar}.

Comparison of the experimental field pattern to the calculated dipolar fields gives remarkable similarity to the case calculated for $\varGamma_3(1)$ structure - the calculated dipolar field directions are close to the experimental values in FIG.~\ref{str:6K}, although the calculated B$_c$ value is relatively small and the calculated local field is 2.7 times larger than the experimental value. Three times larger value of the calculated dipolar field compared to the experimentally determined local field was reported earlier for dipolar field at Ba site in antiferromagnetic YBa$_2$Cu$_3$O$_{6.05}$ Ref.~\cite{Lombardi1996}. The authors ascribed this controversy to the possible effect of delocalization of the copper $d$~-~electron. Dipolar field calculated for $\varGamma_3(2)$ structure is quite different. It is almost confined to the $ab$ plane, with nearly equal B$_a$ and B$_b$ components. Therefore, we can conclude that the NMR data are better consistent with the magnetic structure $\varGamma_3(1)$.\\

\section{Conclusion}
We performed $^{31}$P NMR of the antiferromagnetic square cupola compound \BTCP/ in  the applied magnetic field B = 4.7~T and provided an in-depth overview of the local magnetic environment around the Cu$^{2+}$ cupolas. From the $^{31}$P NMR frequency dependence of the single crystal orientation we successfully determined the principal values of the $^{31}$P magnetic shift tensor and the orientation of eight magnetic tensors in the unit-cell at room temperature and at temperature T~=~6~K. The Knight shift temperature dependence in comparison of that of the bulk magnetic susceptibility enabled to determine the hyperfine field on $^{31}$P nuclei \textit{H}$_{\text{hf}}$~=~7.65(5) kOe/$\mu_B$ for $a\parallel B$ and \textit{H}$_{\text{hf}}$~=~7.40(5) kOe/$\mu_B$ for $c\parallel B$. The temperature dependence of $^{31}$P spin-lattice relaxation resulted in an approximation for the exchange interaction constant between Cu$^{2+}$ ions $J~= ~35K$. $^{31}$P NMR frequency dependence on the single crystal orientation in the antiferromagnetic state gave a clear picture of local magnetic fields at  $^{31}$P ions. The static magnetic field at every phosphorus was determined as $B_\text{int}~=~35.6$~mT. Experimental configuration of the local field was compared to the calculated dipolar field for several magnetic arrangement of the copper magnetic moments. We found that the magnetic structure $\varGamma_3(1)$ determined by the previous neutron diffraction studies~\cite{Kimura2016magneto,Babkevich2017} is most consistent with the NMR data. 

\section*{Acknowledgemnts}
This research was supported by the Estonian Science Council Grants IUT23-7 and PRG4, the European Regional Development Fund Grant TK134, JSPS KAKENHI Grant Numbers JP17H01143, JP19H05823 and JP19H01847, and by the MEXT Leading Initiative for Excellent Young Researchers (LEADER).

\end{document}